\documentstyle[aps,prl,epsf]{revtex}

\begin{document}
\newcommand\beq{\begin{equation}}
\newcommand\eeq{\end{equation}}
\newcommand\bea{\begin{eqnarray}}
\newcommand\eea{\end{eqnarray}}

\def\eps{\epsilon}
\newcommand{\ket}[1]{| #1 \rangle}
\newcommand{\bra}[1]{\langle #1 |}
\newcommand{\braket}[2]{\langle #1 | #2 \rangle}
\newcommand{\proj}[1]{| #1\rangle\!\langle #1 |}
\newcommand{\ba}{\begin{array}}
\newcommand{\ea}{\end{array}}
\newtheorem{theo}{Theorem}
\newtheorem{defi}{Definition}
\newtheorem{lem}{Lemma}
\newtheorem{exam}{Example}
\newtheorem{prop}{Property}
\newtheorem{propo}{Proposition}
\newtheorem{cor}{Corollary}
\newtheorem{conj}{Conjecture}

\twocolumn[\hsize\textwidth\columnwidth\hsize\csname
@twocolumnfalse\endcsname

\author{Barbara M. Terhal$^1$ and Karl Gerd H. Vollbrecht$^2$}

\title{The Entanglement of Formation for Isotropic States}

\address{\vspace*{1.2ex}
            \hspace*{0.5ex}{$^1$ IBM Watson Research Center,
P.O. Box 218, Yorktown Heights, NY 10598, USA; $^2$ Institut f\"ur
Mathematische Physik, TU Braunschweig, Mendelssohnstr.3, 38106
Braunschweig, Germany}\\ 
Email: {\tt terhal@watson.ibm.com}, {\tt k.vollbrecht@tu-bs.de}}

\date{\today}

\maketitle
\begin{abstract}
We give an explicit expression for the entanglement of formation
for isotropic density matrices in arbitrary dimensions in terms of
the convex hull of a simple function. For two qutrit isotropic states we
determine the convex hull and we have strong evidence for its exact form 
for arbitrary dimension. Unlike for two qubits, the entanglement of
formation for two qutrits or more is found to be a nonanalytic
function of the maximally entangled fraction in the regime where
the density matrix is entangled.
\end{abstract}

% close bracket for pretty quant-ph mode
]

One of the main goals in quantum information theory is to develop a
theory
of entanglement. A cornerstone of this theory will be a good
measure of bipartite entanglement. Such a measure must obey the
essential property that the entanglement of a bipartite density matrix
$\rho$
which is shared by Alice and Bob cannot increase, on average, under
local
quantum operations and classical communication ($LO+CC$) between Alice and Bob. In
this way, the entanglement captures the truly quantum correlations in
a bipartite density matrix. For pure bipartite states a good measure of
entanglement
has been found, it is the following quantity:
\beq
E(\ket{\psi}\bra{\psi})=S({\rm Tr}_B(\ket{\psi}\bra{\psi})),
\eeq
where $S(\rho)$ is the von Neumann entropy of $\rho$, i.e.
$S(\rho)=-{\rm Tr}\, \rho \log \rho$ and ${\rm
Tr}_B(\ket{\psi}\bra{\psi})$ is the reduced density
matrix that we obtain by tracing out over Bob's quantum system. This
measure $E$ is unique \cite{vidalem,emeas} if one requires the entanglement to obey a set of natural
properties, such as convexity, non-increase under local measurements,
asymptotic continuity, partial additivity and normalization. Moreover, $E$ is a measure of the asymptotic entanglement costs \cite{bbps} of making the state $\ket{\psi}$ out of a canonical set of states, which we can choose to be EPR singlets
$\frac{1}{\sqrt{2}}(\ket{01}-\ket{10})$, which have $E=1$. This process
is reversible, in the
sense that one can concentrate \cite{bbps} a set of $n$ states
$\ket{\psi}$
with entanglement $E$ to a smaller set $m=E n$ EPR singlets.

The situation for mixed states is much more complex. In Ref. \cite{bdsw}
a first measure of mixed state entanglement, called the entanglement
of formation, was introduced. This measure is a candidate for measuring the
asymptotic costs of making the density matrix out of a supply of EPR singlets.
There are no mixed density matrices for which this statement has been proved,
but neither have counterexamples been found so far. The search for a
possible discrepancy between the entanglement of
formation and the asymptotic entanglement costs is
hampered by the fact that we know the entanglement of formation only for
two qubit systems; Wootters \cite{woot} found an analytic expression for the
entanglement of formation for all two qubit density matrices.

In this Letter we present the first calculation of the entanglement of
formation of a class of density matrices in dimensions higher than
${\bf C}^2 \otimes {\bf C}^2$. We explicitly determine the entanglement
of formation for two qutrit density matrices in this class and we find
an expression in arbitrary dimension in terms of the convex hull of a simple function. We conjecture the explicit form of this convex hull, which can be easily
verified in a given dimension. Surprisingly,
the entanglement of formation is found to be a nonanalytic function of
the parameter characterizing the class of states that we consider.
%Knowledge of the entanglement of formation is a first step towards
%solving one of the important open questions in quantum information theory,
%which is the physical interpretation of the entanglement of formation.

Let us start by recalling the definition of the entanglement of formation.
Let ${\cal E}_{\rho}=\{p_i,\ket{\psi_i}\}$ be an ensemble of pure states
which form a decomposition of $\rho=\sum_i p_i \ket{\psi_i} \bra{\psi_i}$. The
entanglement of
formation for mixed states $\rho$ is defined as
\beq
E(\rho)=\min_{{\cal E}=\{p_i,\ket{\psi_i}\}} \sum_i p_i
E(\ket{\psi_i}\bra{\psi_i}).
\label{defe}
\eeq

In this Letter we will consider the class of density matrices, sometimes
called isotropic density matrices, which are convex mixtures of a
maximally
entangled state and the maximally mixed state:
\beq
\rho_{F}=\frac{1-F}{d^2-1}\left({\bf
1}-\ket{\Psi^+}\bra{\Psi^+}\right)+
F \ket{\Psi^+}\bra{\Psi^+},
\label{defwerner}
\eeq
for $0\leq F \leq 1$ and $\ket{\Psi^+}=\frac{1}{\sqrt{d}} \sum_{i=1}^d \ket{ii}$. For $F \leq 1/d$
these density matrices are separable \cite{filterhor}. The entanglement of formation for states with $d=2$ is equal to \cite{bdsw}
\beq
H_2(\mu), \;\; \mu=\frac{1}{2}+\sqrt{F(1-F)},
\eeq
where $H_2(.)$ is the binary entropy function. The states $\rho_F$ have 
the important property \cite{filterhor} that they are invariant under the operation $U \otimes U^*$ for
any unitary transformation $U$. The $LO+CC$ ``twirling''
superoperator ${\cal S}^{U \otimes U^*}$ is defined as
\beq
{\cal S}^{U \otimes U^*}(\rho)=\frac{1}{Vol(U)}\int \, dU\, U \otimes
U^* \rho\, U^{\dagger} \otimes {U^*}^{\dagger}.
\label{defS}
\eeq

In Ref. \cite{terhalsrank} the Schmidt number of the isotropic states
was determined. Instead of making an isotropic state out of
a set of maximally entangled states, we ask how to construct an isotropic
state with a given $F$ out of some state characterized by a Schmidt
vector $\vec{\mu}$.
%The action of ${\cal S}^{U \otimes U^*}$ on matrix elements is the following:
%\bea
%i \neq j,\;\;\ket{ii}\bra{jj} \rightarrow
%\frac{1}{d}P_+-\frac{1}{d(d^2-1)}({\bf 1}-P_+), \nonumber \\
%\ket{ii}\bra{ii} \rightarrow \frac{1}{d}P_++\frac{1-1/d}{d^2-1}({\bf
%1}-P_+),\nonumber \\
%\ket{ij}\bra{ij} \rightarrow \frac{1}{d^2-1}({\bf 1}-P_+),
%\label{twirlrules}
%\eea
%where $P_+$ is the projector onto the state $\ket{\Psi^+}$. All other matrix e%lements vanish under the action of twirling.
So, let us take an arbitrary initial pure
state $\ket{\psi}=\sum_{i=1}^d \sqrt{\mu_i} \ket{a_i,b_i}$ and consider the
effect of twirling. We can write $\ket{\psi}=U_A \otimes U_B \sum_i \sqrt{\mu_i} \ket{i,i}$ and
thus
\bea
{\cal S}^{U \otimes U^*}\left(\sum_{i,j} \sqrt{\mu_i \mu_j}
\ket{a_i,b_i}\bra{a_j,b_j}\right)= \nonumber \\
{\cal S}^{U \otimes U^*}(({\bf 1} \otimes V) \sum_{i,j} \sqrt{\mu_i \mu_j} \ket{i,i}\bra{j,j} ({\bf 1} \otimes V^{\dagger})),
\eea
where $V=U_A^T U_B$. We define $v_{ij}=\bra{i}\, V\, \ket{j}$. The twirled
state becomes
\beq
{\cal S}^{U \otimes U^*}(\ket{\psi}\bra{\psi})=\frac{|\sum_i \nu_i|^2}{d} P_++\frac{1-|\sum_i \nu_i|^2/d}{d^2-1}({\bf
1}-P_+),
\label{restwirl}
\eeq
where $\nu_i=\sqrt{\mu_i} v_{ii}$ and $P_+=\ket{\Psi^+}\bra{\Psi^+}$. When we choose $V={\bf 1}$ we find the density matrix $\rho_F$ at
$F=[\sum_i \sqrt{\mu_i}]^2/d$. For general $V$ one can bound
\beq
|\sum_i \nu_i|^2 \leq [\sum_i |\nu_i|]^2
\leq [\sum_i \sqrt{\mu_i}]^2,
\label{lammu}
\eeq
since $|v_{ii}| \leq 1$ for all $i$. Thus the largest value for $F$ is
obtained by choosing the initial state $\sum_{i=1}^d \sqrt{\mu_i}
\ket{ii}$.

The use of symmetry makes it possible to give a simplified expression for
the entanglement of formation for isotropic states: \\
{\em Lemma 1} The entanglement of formation for isotropic states
in ${\bf C}^d \otimes {\bf C}^d$ ($d \geq 2$) for $F \in (1/d,1]$ is
given by
\beq
E(\rho_F)=co(R(F)),
\eeq
where $co(g)$ denotes the convex hull of the function $g$ and $R(F)$ is defined as
\beq
R(F)=\min_{\vec{\mu}}
\left\{H(\vec{\mu})\,|\,F=[\sum_{i=1}^d
\sqrt{\mu_i}]^2/d\right\},
\label{expr_eform}
\eeq
and $\vec{\mu}$ is a Schmidt vector. \\
{\em Proof} Assume that there exists an optimal decomposition of
$\rho_F$
formed by the ensemble $\{p_i,\ket{\psi_i(\vec{\mu}^i)}\}$, where
$\vec{\mu}^i$ denotes the Schmidt vector of the state $\ket{\psi_i}$. By twirling the l.h.s. and r.h.s. of the equation $\rho_F=\sum p_i
\ket{\psi_i}\bra{\psi_i}$
we obtain that $\rho_F=\sum_i p_i \rho_{F_i}=\rho_{\sum_i p_i F_i}$
where
\beq
F_i(V_i,\vec{\mu}^i)=\frac{|\sum_k v_{kk} \sqrt{\mu_k^i}|^2}{d},
\eeq
as in Eq. (\ref{restwirl}). Since the decomposition is optimal, each Schmidt vector $\vec{\mu}^i$ has minimal entropy under this constraint.
Consider the function
\beq
R_{V}(F)=\min_{\vec{\mu}}\{H(\vec{\mu})\,|\,F=|\sum_{i=1}^d
v_{ii}\sqrt{\mu_i}|^2/d\}.
\eeq
An optimal decomposition of $\rho_F$ is a convex combination of pure states each of which corresponds to a certain $F$ under twirling.
Thus the entanglement of formation $E(\rho_F)$ can be obtained
by taking the convex hull of the functions $co(R_V(F))$. We can make an additional
simplification.
Eq. (\ref{lammu}) implies
that $R_V(F)=R_{\bf 1}(F') \equiv R(F')$ where $F' \geq F$ for every $V$. Thus
instead of taking the convex hull of all functions $co(R_V(F))$, we can
take the convex hull of function $R_+(F)=\min_x \{R(x)\,|\, x \geq F\}$.
In Lemma 2 we will determine $R(F)$ and it is not hard to show that
$R(F)$ is a monotonically increasing function of $F$. It follows then that 
$R_+(F)=R(F)$ and $co(R_+(F)=co(R(F))$. $\Box$

We now determine the function $R(F)$ defined in Eq.
(\ref{expr_eform}). Since all the equations are symmetric in
$\mu_i$, we can restrict ourselves to solutions which satisfy
$\mu_1\geq \mu_2\geq \dots \geq \mu_d$. With the method of Lagrange
multipliers we get a necessary condition for the minimum \beq
-1-\log{\mu_i}+\Lambda_1+\frac{\Lambda_2}{2}\mu_i^{-\frac{1}{2}}=0,
\eeq where $\Lambda_1,\Lambda_2$ denote the Lagrange multipliers.
For fixed $\Lambda_1,\Lambda_2$ this determines the whole set
$\{\mu_i\}$. Setting $\mu_i=\frac{1}{q_i^2}$ we obtain an
expression of the form $\log{q_i}=A q_i +B$ where $A,B$ only
depend on $\Lambda_1,\Lambda_2$. Since a convex and a concave
function cross each other in at most two points, this equation
has maximally two possible nonzero solutions for $q_i$. Therefore
all Schmidt vectors $\vec{\mu}$ that are possible candidates for
the minimum have to satisfy the condition $\mu_i
\in\{\gamma,\delta,0\}$. Let $n$ be the number of entries where
$\mu_i=\gamma$ and $m$ the number of entries where $\mu_i=\delta$.
The minimization problem has been reduced considerably: For fixed
$n , m$, $n+m \leq d$, we minimize the function \beq n h(\gamma)+ m
h(\delta)\label{minfunc} \;,\eeq where $h(x)=-x \log x$, under the
constraints \beq \{n \gamma + m \delta=1 \label{norm} \;,
n\sqrt{\gamma}+m \sqrt\delta=\sqrt{d F}\} \;,\eeq
The constraints give rise to a quadratic equation in
$\sqrt{\gamma}$ which provides two possible solutions for $\gamma$
for every choice of $n,m$: \beq \gamma^{\pm}_{nm}(F)=\left(
\frac{\sqrt{d F}n \pm \sqrt{m n (m+n-d F)}}{n(n+m)}\right)^2
\label{lambda}\;. \eeq With the first constraint we get the
corresponding $\delta^\pm_{nm}(F)=(1-n \gamma^\pm_{nm}(F))/m$.
Since $\gamma^-_{mn}=\delta^+_{nm}$, the function in Eq.
(\ref{minfunc}) takes the same value for $\gamma^+_{nm}$ and
$\gamma^-_{mn}$. Therefore we can restrict ourselves to the
solutions $\gamma_{nm}:=\gamma^+_{nm}$. The pointwise minimum over
all possible choices for $n,m$ of
 \beq
R_{nm}(F)=H_2(n\gamma_{nm})+n\gamma_{nm} \log \frac{n}{m}+\log m,
\eeq
defined on the domain $\frac{n}{d}\leq F \leq \frac{n+m}{d}$,
is the required function $R(F)$. The restriction on the domain
comes from requiring that $\gamma_{nm}$ is a proper solution of
Eq. (\ref{lambda}) which implies that $F \leq \frac{n+m}{d}$. On
the other hand we demand that $\delta_{nm} \geq 0$ which implies
that $F \geq \frac{n}{d}$. In this regime one can verify that
$\gamma_{nm}\geq\delta_{nm}\label{gamdel}$.

When $m=0$, $\gamma$ and $F$ are uniquely determined by the constraints, i.e. $F=\frac{n}{d}$. Since $R_{n0}(\frac{n}{d})=R_{n'm'}(\frac{n}{d})$
for all $n'+m'=n$, we can neglect these cases.

\begin{figure}
\begin{center}
\epsfxsize=8.5cm \epsffile{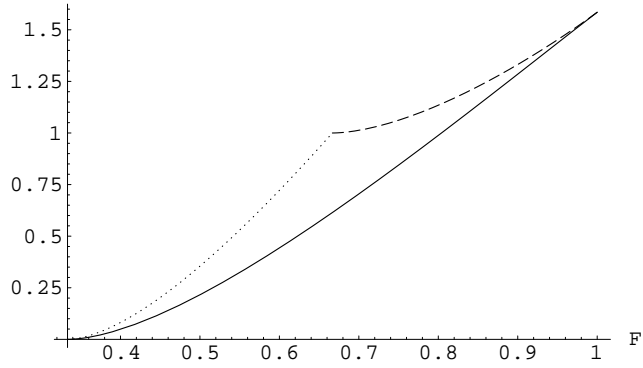}
\caption{
The $R_{nm}(F)$ functions for $d=3$. The solid line is $R_{12}(F)$, which is the minimal one. The dotted and dashed lines are $R_{11}(F)$ and $R_{21}(F)$ respectively.}\label{fig1}
\end{center}
\end{figure}

When $d=3$, what remains is a minimization over the three functions $R_{12}(F)$, $R_{21}(F)$ and $R_{11}(F)$, which are plotted in  Fig. \ref{fig1}.
For $d=3$ we get $R(F)=R_{12}(F)$ \cite{KGHV}.
Thus the optimal vector $\vec{\mu}$ is always of the form

\beq
\vec{\mu}=\{\gamma, \delta, \delta\} \;,\label{looklike}
\eeq
satisfying $\gamma \geq \delta$.

The case $d=3$ is the important one, since it turns out that we can relate all the higher dimensional minima to $d=3$ and prove that \\
{\em Lemma 2} For $d \geq 3$ the function $R(F)=R_{1,d-1}(F)$. \\
 {\em Proof} The case $d=3$ is discussed above.
Note that $R_{1,d-1}(1/d)=0$, which is clearly minimal, so we provide
a proof for $F> 1/d $.
Let the minimum be attained in $d > 3$ dimensions by a vector
$\vec{\mu}=\{{\mu_i}\}$. Let us select some subset of the entries of
$\vec{\mu}$, the set $\{\mu_{i_j}\}_{j=1}^{d'}$, where $\sum_{j=1}^{d'}
\mu_{i_j}=k\leq 1$. Since $\vec{\mu}$ is the minimum, it follows
that the set $\{\mu_{i_j}\}_{j=1}^{d'}$ is the minimum when we keep
the other entries of the vector $\vec{\mu}$ fixed. Let $\mu_j':=\frac{\mu_{i_j}}{k}$. The vector $\vec{\mu'}$ is the solution for the minimization of
\beq
\sum_{i=1}^{d'} h(\mu_i' k)=k\left[\sum_{i=1}^{d'}h(\mu_i')\right]+h(k),
\eeq
under the constraints $\sum_{j=1}^{d'}\mu_j'=1$ and $\sum_{j=1}^{d'} \sqrt{\mu_j'} ={\cal C}$ where
\beq
{\cal C}=\sqrt{\frac{dF}{k}}-\frac{1}{\sqrt{k}}\sum_{i|\forall j,i \neq i_j}\sqrt{\mu_i}\,.
\eeq
This last equation can always be written as ${\cal C}=\sqrt{d' F'}$ for some $F'$. Thus the restricted minimization problem is equivalent to a $d'$-dimensional version of the original problem, up to the scaling factor $k$ and the additive term $h(k)$. When $F' \leq \frac{1}{d'}$, we know that solution of this
minimization problem is given by a Schmidt vector $\vec{\mu}'$ which
corresponds to an unentangled state, i.e. it is of the form $\vec{\mu}'=\{1,0,\ldots,0\}$.
Let us choose three arbitrary $\mu_i$ out of the optimal vector $\vec{\mu}$.
When the resulting $F' \leq \frac{1}{3}$, it follows that $\vec{\mu}'=\{1,0,0\}$. When $F' > \frac{1}{3}$, the three entries of $\vec{\mu}'$ have to satisfy
Eq. (\ref{looklike}). So in fact, in both cases they satisfy Eq. (\ref{looklike}).
Suppose now that one entry of $\vec{\mu}$ is equal to zero. Then it follows that
$\vec{\mu}$ cannot have two nonzero entries since this would violate
condition (\ref{looklike}), in other words it must be that $\vec{\mu}=\{1,0,\dots,0\}$.
But this is  a solution for  $F=1/d$. Therefore we get $n+m=d$ for $F>1/d$.
Suppose that $n\geq 2$. We can choose the vector
$\{\gamma,\gamma,\delta\}$ satisfying $\gamma \geq \delta$. Then condition
(\ref{looklike}) implies that $\gamma=\delta$. This implies that all entries of
$\vec{\mu}$ are identical, or $\vec{\mu}=\{\frac{1}{d},\dots,\frac{1}{d} \}$.
This corresponds to a maximally entangled state, which is the
unique solution for $F=1$. Therefore $n=1$ and $m=d-1$.
 $\Box$

%It is interesting to note that this method of relating the higher dimensional
%minimum to the lower dimensional minimum does not work if we try to derive
%the optimal solution solely from the $d=2$, the two qubit, case.

Lemma 1 and Lemma 2 together result in \\
{\em Theorem 1} The entanglement of formation $E(\rho_F)$ for isotropic states in
${\bf C}^d \otimes {\bf C}^d$ ($d \geq 2$) for $F \in (1/d,1)$
is given by
 \beq E(\rho_F)=co(R_{1,d-1}(F)),
\eeq
where
\beq
R_{1,d-1}(F)= H_2(\gamma(F))+(1-\gamma(F))\log{(d-1)}, \label{expr}
 \eeq
with
\beq
\gamma(F)=\frac{1}{d}\left( \sqrt{F} +
\sqrt{(d-1)(1-F)}\right)^2\,.
\eeq

\begin{figure}
\begin{center}
\epsfxsize=8.5cm \epsffile{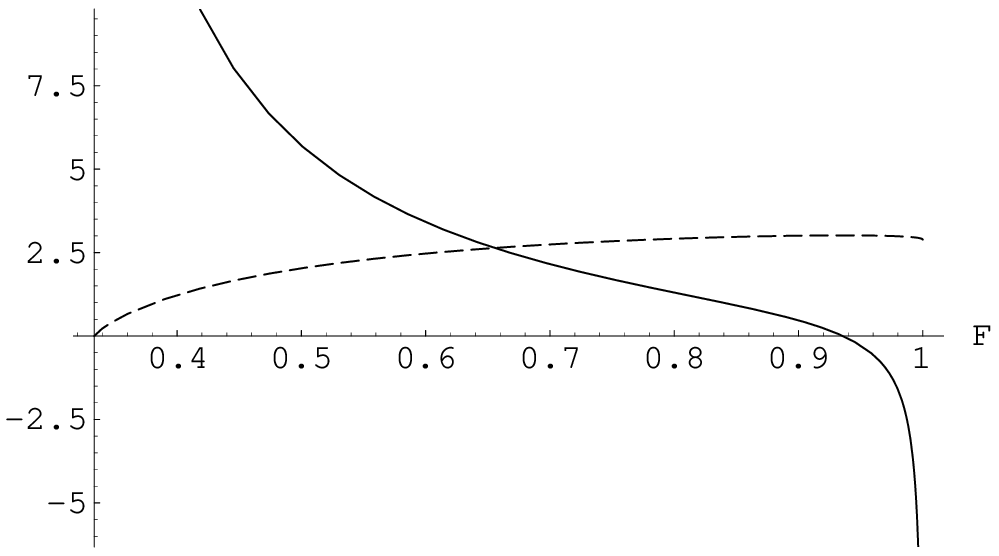}
\caption{
The first and second derivative of $R_{1,d-1}(F)$ for $d=3$. The solid line is
the second derivative, which is going to $-\infty$ for $F=1$.
The dashed line is the first derivative.
}\label{fig2}
\end{center}
\end{figure}

For $d=3$ the first and second derivative of the function $R_{12}(F)$ are
plotted in Fig. \ref{fig2}. The figure shows that the function $R_{12}(F)$ is not convex near
$F=1$; its second derivative is not positive. In order to determine
$co(R_{12}(F))$ for $d=3$ we solve the following
equations. Let $E_{line}(F)=a F+\log 3 -a$ be the line crossing
through the point $(1,\log 3)$. We solve (1) $E_{line}(F)=R_{12}(F)$ and
(2) $\frac{d E_{line}}{dF}=a=\frac{d R_{12}}{dF}$ for $a$ and $F$. Figure \ref{fig2}
indicates that $R_{12}(F)$ is monotonically increasing and that there is only 
one region where $R_{12}(F)$ is not convex, namely near $F=1$. Therefore the 
solution to the equations will be unique: we find that $F=8/9$ and $a=3$.
For higher dimensions, we conjecture, based on examining these two equations,
that the entanglement of formation in ${\bf C}^d \otimes {\bf C}^d$ is
given by
\beq
E(\rho_F)=\left\{\ba{ll} 0, & F \leq \frac{1}{d}, \\
                R_{1,d-1}(F), & F \in
\left(\frac{1}{d},\frac{4(d-1)}{d^2}\right), \\
\frac{d\log(d-1)}{d-2}(F-1)+\log d, & F \in [\frac{4(d-1)}{d^2},1].
\ea\right.
\eeq

The correctness of this solution can easily be verified for a given $d$
by plotting the function $R_{1,d-1}(F)$ and its second derivative and noting the
convex hull of $R_{1,d-1}(F)$ is obtained by calculating where $R_{1,d-1}(F)$ meets
the line going through the point $(F=1,E=\log d)$ and the tangent of
$R'_{1,d-1}(F)$ equals the slope of this line.

It is surprising to find that $E(\rho_F)$ is nonanalytic in the
region where $\rho_F$ is an entangled density matrix. Another
feature of our solution is that for, say, $d=3$ and $F > 8/9$ an
optimal decomposition of $\rho_F$ is not one in which every pure
state has an equal amount of entanglement. Indeed, the optimal
decomposition that we find, is a mixture of the maximally
entangled state and the ensemble of states $\ket{\psi}$ obtained 
by twirling, each of which has entanglement $E= -1/3+\log 3$. Since every
state in the optimal decomposition of $\rho_F$ has, under twirling, a value
of entanglement on $R_{1,d-1}(F)$, every optimal decomposition of
$\rho_F$ for $d=3$ in the range $F > 8/9$ will be a mixture of the
maximally entangled state and some less entangled states. This
is in contrast with optimal decomposition for $E$ for two qubits.
For $F > 8/9$ more than $d^2=9$ pure states must be used in the 
optimal decomposition of $\rho_F$. We make
$\rho_F$ from a maximally entangled state and the state
$\rho_{F=8/9}$ which has rank 9, and thus needs at least 9 states
in its optimal decomposition. In total, this gives 10 states. For $F> 8/9$ 
there is no optimal decomposition with fewer states: one always has to 
mix in the maximally entangled state with some probability. The 
remaining state $\rho_F'$ either has rank 9 (like $\rho_{F=8/9}$) or a lower 
rank. If it has a lower rank, it must be separable, which would imply
that the optimal decomposition is made from mixing a separable state 
with a maximally entangled state which we know to be false.
This is the first example of an entangled state for which it is proved that 
the number of pure states in the optimal decomposition exceeds the rank
of the state (see Ref. \cite{barely} for separable states with this property).

Crucial in our method is the invariance of the isotropic states under
a symmetry group of local operations. A result similar to Lemma 2 will
hold for example for the class of Werner states  \cite{werner:lhv} which are
invariant under the transformation $U \otimes U$ for all $U \in U(d)$.
Let ${\cal S}^{U \otimes U}$ be defined as in Eq. (\ref{defS}), but with
omission of the complex conjugation. The Werner states $\rho_p^W$ are characterized by a single parameter $p$. One can prove that $E(\rho_p^W)=co(R(p))$
where
\beq
R(p)=\min\left\{E(\ket{\psi}\bra{\psi})\;|\; {\cal S}^{U \otimes U}(\ket{\psi}\bra{\psi})=\rho_p^W\right\}.
\eeq
It may thus be possible to carry out a similar analysis as was done here for
the Werner states. In a further generalization one could consider the 
entanglement of formation for $g \otimes g$ or $g \otimes g^*$ invariant 
states where $g \in G$ and $G$ is a subgroup of $U(d)$.

{\bf Acknowledgments}: BMT would like to thank David DiVincenzo, Julia Kempe, John Smolin and Armin Uhlmann for interesting discussions. BMT acknowledges support of the ARO under contract number DAAG-55-98-C-0041. KGHV would like to thank R.F. Werner for discussions. KGHV is supported by Deutsche Forschungs Gemeinschaft (DFG).

\end{document}